\documentstyle[aps,prl,epsf,multicol]{revtex}
\begin{document}
\draft


\title{Flux penetration in slab shaped Type--I superconductors}

\author{Hemant Bokil and Onuttom Narayan}
\address{Department of Physics, University of California, Santa Cruz, CA 95064}

\date{\today}

\maketitle

\begin{abstract}

We study the problem of flux penetration into type--I
superconductors with a high demagnetization factor (slab geometry). 
Assuming that the interface between the normal and superconducting
regions is sharp, that flux diffuses rapidly in the normal
regions, and that thermal effects are negligible, we analyze 
the process by which flux
invades the sample as the applied field is increased slowly from zero.
We find that flux does not penetrate gradually. Rather there is an
instability in the process and the flux penetrates from the boundary
in a series of bursts, accompanied by the formation of isolated droplets 
of the normal phase, leading to a multiply connected flux domain
structure similar to that seen in experiments.

\end{abstract}
\pacs{74.55.+h, 05.70.Ln, 75.60.-d}
\begin{multicols}{2}
\narrowtext

When a type--I superconductor is placed in a magnetic field less than
the bulk upper critical field it exhibits a phase with interpenetrating
domains of the normal and superconducting state called the intermediate state. 
Despite over fifty years of work, a complete description of the
physics of this phenomenon has been elusive.
The earliest theoretical work on this problem goes
back to Landau ~\cite{landau} who studied the problem of the equilibrium
configuration of an infinite superconducting slab of thickness $d$
placed in an applied magnetic field $B_{app}$. 
When $d$ is much larger than the penetration depth
$\lambda$ and the coherence length $\xi$, 
he showed that the Meissner
state is unstable to a 
configuration of alternating superconducting and normal laminae.
Other configurations of flux domains, various sample geometries, etc.
have been studied both experimentally and theoretically~\cite{parks},~\cite{huebener}(and references therein).

It has been clear since the work of Landau that these
structures arise from a competition between the magnetic field energy,
the condensation energy of the superconducting
regions, and the surface energy of the interface between the normal and
superconducting regions. 
However, the striking fact about most of the experiments is that
the regular structures envisioned by Landau are
rarely seen. Instead one sees complicated patterns
which are strongly
dependent on the temperature, disorder and field history of the sample.

In the last few years considerable progress has been made in understanding
a related problem: flux penetration into (or expulsion
from) long cylindrical samples oriented parallel to the magnetic 
field ~\cite{dorsey1}.
Although interesting flux patterns are seen in these samples
as {\it transients\/} in going from a normal to a superconducting state,
a steady--state intermediate state, which arises due to demagnetization effects,
is not seen. Very recently,
Goldstein, Jackson and Dorsey (hereafter referred to as GJD) ~\cite
{goldstein} attempted to analyze the
influence of demagnetization on the formation of flux structures for
slab--like
type--I superconductors. They assumed that transient currents in the 
normal domains decay very fast (which amounts to the normal state
conductivity being zero), 
and argued that one can then think of  
the dynamics as a simple gradient descent of the free--energy. When
the coherence length $\xi$ is small, the interface between normal
and superconducting regions is approximately sharp. 
The free--energy for a sample placed in an applied 
magnetic field $B_{app}$ can then be written as

\begin{equation}
\label{free_en}
{\cal F}(B_{app},\Delta) = {\cal F}_B + {\cal F}_c + {\cal F}_s.
\end{equation}

Here the first term denotes the magnetic free energy, the 
second term the condensation energy which is $ - H_c^2/8\pi$ per unit
volume in the superconducting regions, and the
third the interfacial energy which is $H_c^2 A\Delta/8\pi$. $\Delta$ is the surface energy
parameter which is of the order of the coherence length $\xi$, $H_c$ is
the bulk upper critical field and $A$ is the area of the interfaces
between the normal and superconducting regions.

GJD ~\cite{goldstein}
{\it assumed} that (in addition to a bulk term) the magnetic part of this energy could be 
written as a long--range interaction
between current loops localized on the interfaces between the normal and
superconducting regions.
This made the problem very similar to the problem of the
dynamics of two--dimensional ferrofluid droplets in a magnetic field
~\cite{ferro}. In the 
ferrofluid case it is known ~\cite {ferro} that regular shapes 
evolve {\it continuously\/} into labyrinthine patterns 
when the applied field is 
increased adiabatically. In the case of superconductors,
GJD showed that a circular flux droplet in a sea of
superconducting material (with an area much larger than the equilibrium
area) changes into a many armed
structure with three--fold coordinated nodes.
Such convoluted structures are indeed
seen in some experiments~\cite{parks}~\cite{huebener}. They also
calculated the equilibrium periodicity for a laminar structure and got
results that are numerically close to Landau's. 
While this would suggest that their free--energy captures the
physics of this situation, 
GJD ~\cite{goldstein} were 
unable to derive it from the more basic Ginzburg--Landau
description. 

As shown below, a careful analysis of the magnetic 
free energy yields an interaction between superconducting
domains that {\it cannot\/} be expressed in terms of 
current carrying loops (at least for low fields), and results in qualitatively
different behavior. (Late stages of flux invasion, when the supercondcting
domains are tall and thin, are discussed later.) It is not difficult to see why superconductors
and ferrofluids are so different. When a thin layer of a ferrofluid
of thickness $d$ and a characteristic transverse linear extent of $L$
is placed in an applied magnetic field $B_{app}$, to leading order in $d/L$
the
$B$ field inside (and outside) the ferrofluid is equal to $B_{app}$. 
This uniform field induces a uniform magnetization within the sample,
which (using the equation $\nabla\times B={4\pi\over c}j_{ext}+
4\pi\nabla\times M$)
gives rise to a small fringing $B$ field near the sample edges.
(The fringing field in turn induces a higher order non--uniform 
magnetization.) To leading order in $d/L$
the induced currents consist of a ribbon flowing around the boundary
of the ferrofluid, causing a long--range interaction between 
different parts of the boundary. On the other hand, for a thin 
superconductor in the intermediate phase, $B$ is not even approximately
equal to $B_{app}$ near the superconducting regions. Outside the
sample, just above or below a superconducting domain, $B$ is {\it parallel\/}
to the surface, while inside the domain $B$ is zero. In 
addition to ribbons of current 
along the side walls of the superconducting regions, 
there are also large current sheets on the top and bottom surfaces,
dominating the inter--domain interaction. Thus 
while the basic GJD idea of long--range interactions destabilizing 
regular patterns is correct, the actual description of the experimental
patterns is more complicated. The numerics 
we report in this paper show flux invading in bursts, pinching 
off from the boundaries to form droplets. This is qualitatively
different from the continuous evolution of GJD ~\cite{goldstein}.

\begin{figure}
\centerline{
\epsfxsize\columnwidth\epsfbox{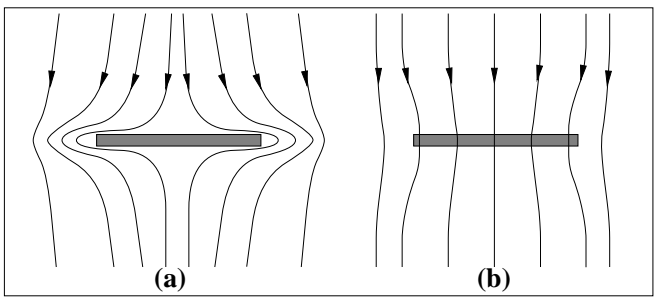}}
\vskip 0.5truecm
\caption{$B$-field for a slab shaped (a) superconducting domain
(b) ferrofluid in a vertical applied field. The distortion
of the field is small only for (b).}
\label{fig1}
\end{figure}

We now consider ${\cal F}_B$ in detail.
For an arbitrary sample placed in an applied magnetic field $B_{app}$,
the magnetic free--energy is~\cite{degennes}
${\cal F}_B = {1\over{8\pi}}\int d^3 x (B^2 -2 \vec B. \vec H)$.
Here $B$ is the magnetic induction, $H$ is the magnetic field
and the integral is over all of three
dimensional space. It is convenient to cast this equation in a 
slightly different form: since $\nabla\times \vec H=\nabla\times \vec B_{app}$,
and $B$ is a transverse field, $\vec H$ can be replaced
by $\vec B_{app}$ above. Adding the $B$--independent term
$B_{app} ^2/8\pi$ to the free energy density yields 

\begin{equation}
\label{free_mag3}
{\cal F}_B = {1\over{8\pi}}\int d^3 x (B - B_{app})^2
\end{equation}

Thus in order to evaluate ${\cal F}_B$, one has to 
find the magnetic induction $B$ for a given applied field
$B_{app}$. For the superconducting regions, where $B = 0$, 
the contribution to ${\cal F}_B$ is simple.
The non--trivial
part of the calculation consists of determining the magnetic 
field in the normal domains and in the space outside. 
Outside the superconducting regions, one can define a magnetic 
scalar potential $\phi$ by 
$B - B_{app} = \bigtriangledown \phi$, satisfying $\nabla^2\phi=0$.
(We assume that the applied magnetic field is not changed rapidly,
so that transient currents in the normal regions can be ignored.)
Given any configuration of superconducting and
normal regions we then have to solve Laplace's equation outside the
superconducting regions. Since the normal component of $B$ is zero
at the boundaries, for a flat superconducting slab in a vertical 
field $B_{app}$ pointing upwards the boundary conditions are 
(i) $\partial_n \phi = \pm B_{app}$ on the top and bottom surfaces of the
superconducting regions and
(ii) $\partial_n \phi = 0$ on the interfaces between the normal and
superconducting regions, where $\partial_n$ is the normal
derivative.  (We have assumed that the interfaces are
vertical, ignoring fanning out of flux
domains at the top and bottom surfaces.) 

At this stage, we use the quasi two--dimensional
nature of the problem to reduce its computational complexity. For
a thin sample, most of ${\cal F}_B$,
${1\over{8\pi}}\int d^3 x (\nabla\phi)^2,$ is stored 
outside the sample (and in the normal regions). Integrating
by parts transforms this expression into ${1\over{8\pi}}\int_S ds
\phi\partial_n\phi$, where the integral runs over the top and bottom
surfaces of the superconducting regions. (On the side walls, the 
normal derivative of $\phi$ is zero.)
In evaluating the surface integral, for a thin sample the 
top and bottom surfaces of the superconducting regions can be 
treated as approximately coincident, lying in the $z=0$ plane. 
By symmetry, $\phi(z=0)=0$, except in the superconducting regions
where $\phi(z=0^+)=-\phi(z=0^-)\neq 0$. Using the boundary condition
$\partial_z\phi(z=0^\pm)= - B_{app}$ for the superconducting regions
yields 

\begin{equation}
{\cal F}_B={{B_{app}}\over{4\pi}}\int d\vec r \phi(z=0^+,\vec r)
\end{equation}

where $\vec r$ is a two--dimensional vector.
The integral is performed only over the superconducting regions.
To obtain $\phi(z=0^+)$ we solve the inverse 
problem: if $\phi(z=0^+)$ were known, $\partial_n\phi(z=0^+)$ could
be found by solving Laplace's equation with Dirichlet boundary
conditions. Using $\partial_z\phi(z=0^+)=-B_{app}$ yields the 
condition 
\begin{equation}
B_{app} =-\partial_z\phi(z=0^+,\vec r)=\int d\vec r^\prime K(\vec r-\vec r^\prime)
\phi(z=0^+,\vec r^\prime).\label{selfcons}
\end{equation}
The integral on the right hand side is over the superconducting 
regions.
$K$ is given by
\begin{equation}
K(\vec r-\vec r^\prime)=\lim_{z\rightarrow 0}{1\over{2\pi}}
{{2z^2-|\vec r-\vec r^\prime|^2}\over
{[|\vec r-\vec r^\prime|^2+z^2]^{5/2}}}
\end{equation}
whose (two dimensional) Fourier transform is given by $|k|/2\pi$. 
Since $K$ is convoluted with $\phi$, which is a smooth function, it can be 
replaced by 
\begin{equation}
K(\vec r-\vec r^\prime)=-1/[2\pi|\vec r-\vec r^\prime|^3] \qquad |\vec r-\vec r^\prime|
>\epsilon\label{kernel}
\end{equation}
with a compensating $\delta$--function of strength $1/\epsilon$ at the origin, in the 
limit $\epsilon\rightarrow 0$. (For the numerics presented in this paper, the cutoff
$\epsilon$ is effectively the lattice size.) 

For an arbitrary pattern of superconducting and normal regions, 
one can only evaluate ${\cal F}_B$ numerically using the prescription
above. It is not possible to reduce the expression to integrals
over domain boundaries. (Note that it is {\it not\/} possible to solve
Eq.(\ref{selfcons}) trivially by Fourier transforming, since the right hand 
side is only integrated over the superconducting regions.)
The special case of a single circular 
superconducting region can be solved analytically~\cite{jackson}, and 
can be used as a check of the numerics.

We now discuss the numerics.
Since the experimental patterns depend a great deal on the 
field history of the sample, we concentrate on the following 
question:
if one slowly increases the field from zero, how does
flux penetrate the sample?  
We choose a sample with a thickness $d \simeq 12 \Delta$ and linear 
dimension $L = 10 d$ (as stated before $\Delta$ is essentially the 
coherence length $\xi$). In experiments the flux domains branch near
the surfaces for $d>O(800\Delta)$, and type--II behavior is seen for
$d< O(\Delta)$.
Our choice of $d$ avoids both these regimes.
Because $L>>d$,  we can use the two
dimensional formulation of the magnetic boundary value problem to find
${\cal F}_B$.
A typical value for type--I superconductors, $\Delta \simeq 1500 \AA,$
corresponds to $L\simeq 15 \mu m$. Although this is much smaller 
than experimental sample sizes, the qualitative aspects of our results 
should apply to larger samples as well.

We divide the sample into lattices of various sizes; for a $61\times 61$
lattice the lattice constant is approximately $2.4\Delta$. The plaquettes
in the lattice are either superconducting or normal. Apart from the isolated
normal droplets, the superconducting--normal boundary consists of an outer
interface close to the sample edges. There are two basic moves in the 
numerics: changing any one superconducting plaquette on the outer interface
to a normal plaquette, or vice versa. (The dynamics of the droplets are 
different, due to flux conservation, and are discussed later.)
Both possibilities 
need to be allowed for in order allow for droplet formation. A plaquette 
is flipped if there is a force that favors the move; the force is the gradient
of the free energy, which is given by the sum in Eq.(\ref{free_en}). 
The surface tension force (from ${\cal F}_s$)is calculated by smoothing 
the lattice interface and computing the local curvature.

Trying single plaquette moves requires a special treatment of 
surface tension in the dynamics, since flipping a  single plaquette 
corresponds to a small sharp protrusion of the interface, 
with roughly the same forward and lateral extent. 
For a small lattice constant such a move will always
cost a large surface energy compared to the energy gained 
from the magnetic and condensation terms. Thus in the lattice
dynamics whenever the force on a segment of the interface 
favors flipping a plaquette, it is likely that at the 
next time step it will be favorable to revert the plaquette 
to its original state. This is because in a 
lattice approximation the interface is forced to make 
larger excursions than it would like to; it would be better
for a segment of the interface to move forward by only a 
fraction of a plaquette, and then let neighboring segments
catch up. We use a simple prescription to cure this lattice effect:
when a plaquette is flipped, it is not allowed to flip back
at the {\it very next\/} time step, although it can flip back
thereafter. This should give rise to errors in the pattern
only of the order of a lattice constant. 

We start with $B_{app} = 0$, when the whole sample is superconducting.
We raise the field till one plaquette on the boundary becomes normal.
Flipping this plaquette can make it favorable to flip other plaquettes,
in which case we let the system evolve till it reaches a stationary state.
At this point the field is raised again. This process simulates the 
adiabatic increase of the magnetic field that we wish to study.
Isolated droplets have to be handled differently, since the flux in 
them is conserved. In our simulations this constraint is obeyed 
approximately: once a droplet is formed, we move it rigidly in the
direction of the force on it till the force is zero.
At this stage we adjust the number of plaquettes in the droplet so
that the flux in it is as close as possible to its original value.

\begin{figure}
\centerline{\epsfxsize=\columnwidth\epsfbox{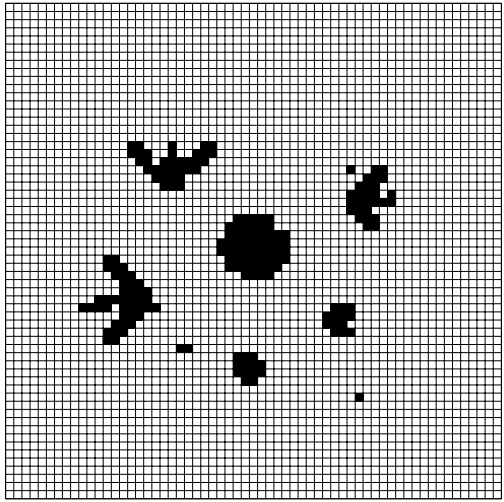}}
\vskip 4truecm
\caption{Droplet state for (a) numerical simulations on 
a $61\times 61$ lattice (b) experiments on mercury 
~\protect\cite{farrell}.
Normal regions are black in (a) and white in (b). The flux
front near the top in (b) is due to finite sample thickness 
~\protect\cite{dummy}}
\label{fig2}
\end{figure}

We now discuss our results.
When the field is raised just above the lower critical field,
one plaquette becomes normal. If the penetration of flux
were gradual, one would expect to have to 
raise the field further for more flux to enter.
Instead we found that at a  field just slightly above the lower
critical field, the flux penetrates a distance into the sample of
the order of twenty times the coherence length before the first droplet
pinches off. Increasing the field further produces similar behaviour:
much of the evolution is in the form of bursts
of magnetic flux penetrating from the boundary which 
then pinch off to form droplets.
This reflects an instability in the process of flux
penetration and is the main result of our work.
In Fig.~\ref{fig2} we show the patterns seen for 
$L = 61$ at a field only moderately above the field of first flux
penetration. We saw similar patterns at comparable fields for
the other lattice sizes.
The droplets form near the boundary of the sample
and then move towards the center of the sample,
leaving a region near the boundary flux--free, similar to what 
is seen in experiments. 
We also saw that the
droplets typically shrink when one puts in flux conservation,
though this observation may not have much
experimental significance in view of the fact that we 
deal with the collective motion of the droplet approximately. 

The formation of droplets is promoted by the discontinuous 
nature of the flux invasion, which causes relatively large amounts of
flux to enter the system at low magnetic fields. A large amount of
flux penetration reduces the magnetic forces which tend to drive 
further flux into the system. It is then possible for a normal region
to find it favorable to revert to being superconducting, which is
how droplets are formed. At later stages when the 
applied field is higher such reversion to a superconducting state
becomes less likely. Further flux coming in from the boundary would
then probably coalesce with the already existing droplets, leading to the 
labyrinthine patterns seen in some experiments, reminiscent of the patterns obtained by
GJD.~\cite{goldstein} However, since the superconducting regions 
become thinner at higher fields, it is not clear whether quasi 
two--dimensional descriptions are valid here. 

Although the existence of
droplets is fairly ubiquitous in the experiments, the actual shape
of the droplets varies rather widely, from compact droplets
in experiments on mercury in the early stages of flux penetration
~\cite{farrell}, to long laminar structures in
experiments on lead ~\cite{kirchner}. In order to explore this
further, one would need to satisfy the constant flux constraint
accurately, as well as construct a continuum description (presumably
analytical) of the dynamics. A continuum analysis
would also be necessary to treat surface
tension exactly.

To conclude, we have developed a description of the problem of flux
penetration into slab shaped Type--I superconductors 
based on the sharp interface approximation. Numerical
simulations on a lattice show that as the applied magnetic field is 
increased, flux penetrates in bursts, forming droplets,  
leading to isolated normal regions.
While the multiply connected nature of the patterns has been emphasized
in the literature ~\cite{huebener}, the 
instability that we have noticed does not seem to have been reported
so far. This instability should be apparent in real time imaging of the
process of flux penetration.

We thank Alan Dorsey, Daniel Fisher, Tanya Kurosky, Carsten Wengel and 
Peter Young for useful 
discussions. HB is supported by the NSF DMR--9411964 and ON in
part by the A.P. Sloan Foundation.

\end{multicols}
\end{document}